\documentclass[prl,aps,twocolumn,floats]{revtex4}

\usepackage{graphics}
\usepackage{amsmath}
\usepackage{dcolumn}

\begin{document}

\title{
{\rm\small\hfill (Chem. Phys. Lett., accepted)}\\
First-principles statistical mechanics approach to step decoration at surfaces}

\author{Yongsheng Zhang and Karsten Reuter}

\affiliation{Fritz-Haber-Institut der Max-Planck-Gesellschaft,
Faradayweg 4-6, D-14195 Berlin, Germany}

\received{14th August 2008}

\begin{abstract}
Using a first-principles parameterized lattice-gas Hamiltonian we study the adsorbate ordering behavior at atomic steps of a Pd(100) surface exposed to an oxygen environment. We identify a wide range of gas-phase conditions comprising near atmospheric pressures and elevated temperatures around 900 K, in which the step is decorated by a characteristic O zigzag arrangement. For catalytic processes like the high-temperature combustion of methane that operate under these conditions our approach thus provides first insight into the structure and composition at a prominent defect on the working surface.
\end{abstract}

\maketitle

\section{Introduction}

As prominent defects at solid surfaces atomic steps are commonly perceived as playing some kind of special, if not decisive role for the surface properties or function in materials science applications. When it comes to the interaction with a reactive environment, steps are viewed as providing particularly active sites which can have a pronounced effect on the catalytic activity or act as nucleation centers for adsorbate-induced morphological transitions like oxide formation or corrosion in our oxygen-rich atmosphere. When aiming to qualify this role at the atomic scale an important first task is to identify the structure and composition at the step edge under realistic gas-phase conditions. On the modeling side, the first-principles atomistic thermodynamics approach \cite{reuter05} has brought considerable progress to this end for ideal low-index surfaces. In its prevalent form, this approach compares the stability of a variety of structural models in contact with a given gas-phase reservoir. Extending it to the study of steps depends then only on the ability to perform the underlying electronic structure calculations for corresponding structural models. For metal surfaces, a viable route to this is to resort to supercell geometries for vicinal surfaces, i.e. to compute periodic arrays of steps that can be suitably cast into a periodic boundary condition framework. Unfortunately, the computational cost connected to such inherently large supercell calculations severely limits the accessible surface unit-cell sizes and total number of structural models that can be computed. This prevents to date a proper exploration of the huge configurational space of possible step structures and allows at best for approximate treatments of entropic effects at elevated temperatures.

In the present study we address these limitations by suitably combining the thermodynamic approach with a first-principles parameterized lattice-gas Hamiltonian (LGH) \cite{reuter05,sanchez84,fahnle05}. The concept is illustrated with the application to the on-surface oxygen adsorption at Pd(100), and there specifically to the decoration of close-packed (111) steps in an oxygen atmosphere. We employ large-scale density-functional theory (DFT) calculations for corresponding low-index and vicinal surfaces to determine the lateral interactions between adsorbates at the (100) terrace sites and at sites close to or at a (111) step. In a first stage the resulting multi-site LGH is then used to generate a large pool of possible structural models with varying on-surface O content. Within the thermodynamic framework the stability of these models is evaluated for a wide range of gas-phase conditions and compared to known more complex O-induced reconstructions at the low-index surface, namely an ultrathin $(\sqrt{5} \times \sqrt{5})R27^{\circ}$ (henceforth simply termed $\sqrt{5}$) surface oxide and thick bulk-like PdO films \cite{zheng02,todorova03,lundgren04}. Intriguingly, this identifies a distinct range of O chemical potentials where a characteristic zigzag step decoration is most stable. Explicitly accounting for the configurational entropy we confirm with grand-canonical Monte Carlo (MC) simulations that this ordered structure is indeed stable up to quite elevated temperatures, in fact up to much higher temperatures than the concurrent ordered adsorbate structure at the Pd(100) terrace. At near atmospheric pressures such temperatures around 900\,K are representative for catalytic applications like the high-temperature combustion of methane \cite{zhu04}, for which our study thus provides first insight into the structure and composition at a prominent defect on the working surface that may serve as a basis for ensuing kinetic studies, e.g. using kinetic Monte Carlo simulations \cite{reuter05}.

\section{Theory}

\begin{figure*}
\scalebox{3.6}{\includegraphics{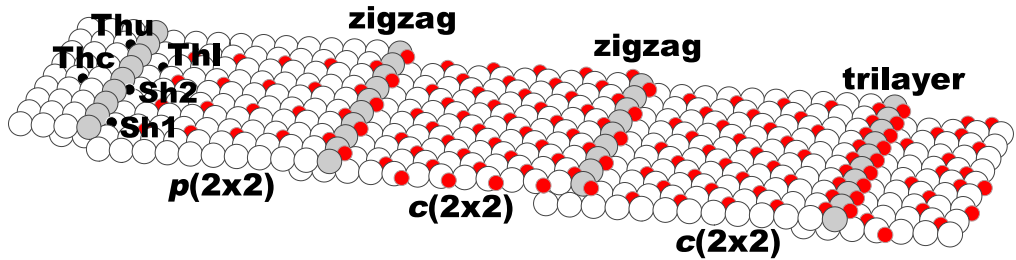}}
\caption{\label{fig1}
Schematic view of a Pd(100) surface with (111) steps. Shown are the five distinct adsorption sites considered in the first-principles lattice-gas Hamiltonian: The fourfold coordinated (100) hollow sites at the terraces (Thc), at the upper (Thu) and lower (Thl) rim of the step, as well as the two threefold (111) hollow sites at the step edge, one with a coordination to two step atoms (Sh2) and one with a coordination to one step atom (Sh1). Additionally represented are four characteristic ordered structures discussed in the text: A $p(2 \times 2)$ and a $c(2 \times 2)$ adsorbate phase at the (100) terrace, a zigzag structure where the step is decorated with O atoms in alternating Thu and Sh2 sites, and a trilayer structure where all Thu and Sh2 are occupied (Pd = large spheres, Pd step atoms = large dark spheres, O = small spheres).}
\end{figure*}

The atomic-level energetics for our first-principles statistical mechanics approach is obtained from DFT calculations, which we perform within the full-potential augmented plane wave (LAPW/APW+lo) framework \cite{wien2k} and using the generalized gradient approximation for the exchange-correlation interaction \cite{perdew96}. In preceding work exactly the same computational setup \cite{basis} was validated and used to systematically investigate the on-surface adsorption of oxygen at Pd(100) \cite{zhang07} and the series of Pd(11$N$) vicinals (with $N=3,5,7$), which exhibit (111) steps and (100) terraces of increasing width \cite{zhang06}. In the limit of low O coverages the energetically most favorable adsorption sites obtained were the fourfold coordinated hollow sites at the extended Pd(100) surface (Thc), as well as the two threefold coordinated hollow sites (Sh1 and Sh2) offered by the (111) step. Furthermore, only the adsorption energetics into the (100) hollow sites directly at the upper and lower rim of the (111) step (Thu and Thl) was noticeably different from that at the corresponding Thc site at the extended Pd(100) surface. We correspondingly focus our present study on these five distinct sites as depicted in Fig. \ref{fig1}, and describe the site-specific adsorption of oxygen atoms on the stepped Pd(100) surface with a lattice-gas model, in which any system state is defined by the occupation of these sites in a lattice. The total free binding energy of any configuration (comprising static and vibrational contributions) is expanded into a sum of discrete lateral interactions between adsorbates at the lattice sites. We consider pair interactions between any combination of sites up to distances that correspond to the third nearest neighor (NN) shell at Pd(100), as well as the most compact many-body interactions in form of trio interactions between any three site combination within a second NN shell distance. Judging from our preceding systematic analysis of lateral interactions at the extended Pd(100) surface \cite{zhang07} we are confident that this truncated LGH covers the ordering behavior at the accuracy level of interest to this study, and we will comment on this accuracy level and steps to further validate the truncation below.

In order to generate a LGH with predictive accuracy the unknown static on-site and lateral interaction energies are parameterized with the DFT binding energies computed for a set of ordered configurations, closely following the recipe detailed for the low-index surface in ref. \cite{zhang07}. In brief, we determine those on-site energies and interactions influenced by the (111) step from large supercell calculations for the Pd(117) vicinal, and those of undisturbed terrace sites from calculations for the low-index Pd(100) surface. In all cases, the ordered structures were fully geometry optimized, which as already detailed in ref. \cite{zhang06} leads to pronounced relaxations of up to 20\% of the first interlayer spacing at the step ridge at high adsorbate coverages. The four on-site energies for the Sh1, Sh2, Thu and Thl sites are accordingly taken from calculations with one O atom inside a Pd(117)-$(1 \times 3)$ surface unit-cell, and the on-site energy for the terrace Thc site from calculations with one O atom inside a Pd(100)-$(3 \times 3)$ surface unit-cell. With distances to the periodic adatom images that are in both cases always larger than 8.37\,{\AA} (corresponding to the sixth NN shell at Pd(100)) we verified that these calculations give a very good approximation to the low-coverage limit. The remaining total of 70 lateral interaction energies are determined by least-squares fitting to DFT binding energies of 102 ordered configurations described in similarly sized supercell geometries.

A first indication for the reliability of the resulting LGH expansion comes from the computed leave-one-out cross validation score \cite{zhang07,shao93,zhang93} of 29\,meV / adatom, which provides a measure for the average accuracy with which the truncated LGH can predict the energetics of adatom configurations not included in the fitting procedure. The consistency of the expansion procedure is furthermore reflected by the similarity of equivalent LGH parameters that can be independently determined by fitting to the low-index or the vicinal DFT data. In all cases such interaction energies differed by less than 20\,meV and we verified that choosing either or the other value did not affect any of the conclusions presented below. Due to the large mass mismatch between O and Pd, the vibrational contribution to the LGH parameters can be approximately considered through its leading term which arises from the change of the O$_2$ gas-phase stretch frequency to the O-substrate stretch mode \cite{zhang07}. Extensive test calculations for various ordered configurations involving all five adsorption sites revealed only a small coverage-dependence of this term, which is correspondingly modeled as constant shifts of the five on-site energies by the zero point energy changes computed in the low-coverage limit.

\section{Results}

In a first stage we exploit the low computational cost to evaluate the thus established first-principles multi-site LGH to extensively explore the huge configuration space and identify low-energy ordered step structures. This is achieved by simulated annealing MC runs in periodic boundary condition simulation cells containing $(60 \times 40)$ (100) surface unit-cells and one (111) step along the shorter cell axis. For fixed and initially random O coverages ranging up to a half filling of all terrace hollow sites $10^8$ MC steps are used to continuously quench from an initial temperature of 2000\,K. Depending on the O content, these simulations unanimously lead to patches of increasing size of four distinct ordered structures that are schematically shown in Fig. \ref{fig1}. At the lowest coverages the O adatoms decorate alternating Thu and Sh2 sites at the (111) step in a characteristic zigzag structure. At increasing coverages domains of the experimentally well characterized $p(2 \times 2)$ overlayer \cite{zheng02} start to form at the (100) sites next to the upper edge of the fully zigzag decorated step. Only once the O content in the simulation cell exceeds the one required to fully cover the entire terrace with the $p(2 \times 2)$ structure, does a further decoration of Thu and Sh2 step sites occur. However, the formation of the resulting periodic step structure which we henceforth term trilayer, cf. Fig. \ref{fig1}, goes then directly hand in hand with the growth of a $c(2 \times 2)$ arrangement \cite{zheng02} at the terrace sites. The first-principles LGH thus correctly reproduces both experimentally characterized ordered terrace structures. As already demonstrated in ref. \cite{zhang07} this holds also for other aspects of the low-coverage phase diagram of the ideal Pd(100) surface like the order-disorder transition, which is better described than by any previous empirical LGH. Without existing experimental data such empirical modeling would furthermore currently be precluded for the adsorbate ensemble near the (111) step, whereas the present first-principles approach clearly predicts two new ordered structures in which the O adatoms decorate the step.

\begin{figure}
\scalebox{0.54}{\includegraphics{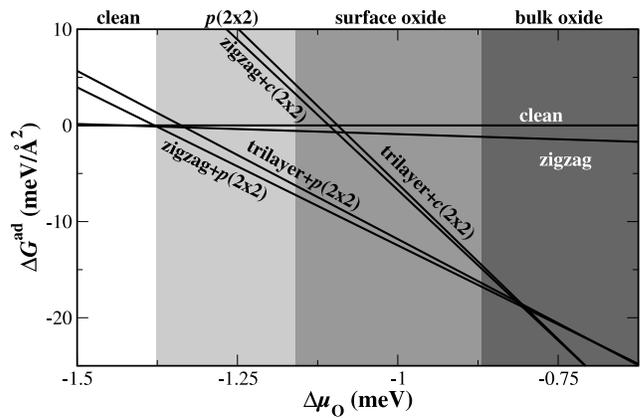}}
\caption{\label{fig2}
Computed Gibbs free energy of adsorption $\Delta G^{\rm ad}$ for O adsorbate structures at a (111) step at Pd(100) and using the clean surface as zero reference. Using the nomenclature defined in Fig. \ref{fig1} the labels refer to the structure at the step and at the neighboring terrace, e.g. zigzag+$p(2 \times 2)$ stands for a zigzag step decoration coexisting with the $p(2 \times 2)$ adsorbate ordering at the terrace. Additionally indicated by differently shaded backgrounds are the stability regions obtained for the ideal Pd(100) surface in previous work \cite{reuter04,rogal07}, namely those of the $p(2 \times 2)$ overlayer, of the $\sqrt{5}$ surface oxide structure and of a thick bulk-like PdO film \cite{density}.}
\end{figure}

For the identified structural models we now account for the effect of a finite gas-phase environment through the atomistic thermodynamics approach \cite{reuter05}. Here, the surface is considered to be in equilibrium with an oxygen gas-phase reservoir characterized by a chemical potential $\Delta \mu_{\rm O}(T,p)$ summarizing the two-dimensional dependence on pressure $p$ and temperature $T$ and using molecular O$_2$ as reference. In its prevalent form this approach neglects the effect of configurational entropy at the surface and then simplifies to approximately computing the Gibbs free energy of adsorption for ordered structural models at any given $\Delta \mu_{\rm O}(T,p)$ from their corresponding binding free energies \cite{reuter01,li03}. This allows to readily compare the stability of most distinct structural models over a wide range of gas-phase conditions, and is thus ideally suited to also assess under which gas-phase conditions the surface structure and composition at the step is appropriately described by our on-surface LGH. We correspondingly employ this approximate approach here not only to obtain a first understanding of the stability of the presented step models, but also to contrast this with the stability ranges of more complex O-rich phases identified previously for the ideal Pd(100) surface \cite{todorova03,lundgren04,reuter04,rogal07,rogal07b}. Specifically, these are the $\sqrt{5}$ surface oxide structure, which corresponds to a sub-nanometer thin film of PdO(101) on the surface \cite{todorova03}, and bulk PdO to represent thick bulk-like oxide films. The results are summarized in Fig. \ref{fig2} and reveal a large stability range of the zigzag decoration of the (111) step, which over most of this range coexists with the $p(2 \times 2)$ adsorbate phase on the (100) terrace. The trilayer arrangement becomes only more favorable, i.e. exhibits a lower $\Delta G^{\rm ad}$, at quite elevated chemical potentials and coexists then directly with the denser $c(2 \times 2)$ overlayer on the terrace. In light of the also shown stability ranges of the oxidic phases at ideal Pd(100) we can therefore identify quite a range of gas-phase conditions ($-1.43$\,eV $< \Delta \mu_{\rm O} < -1.16$\,eV) that are not yet O-rich enough to induce oxide formation and where the zigzag step decoration is not just due to kinetics, but instead represents a thermodynamically stable phase in its own right. In contrast, the elevated $\Delta \mu_{\rm O}(T,p)$ at which the trilayer arrangement at the step becomes more favorable than the zigzag decoration fall well into the stability range of the oxidic surface reconstructions. This suggests an interpretation of the trilayer structure in form of a kinetic precursor in the oxide formation process, and in this respect it is intriguing to note that its geometric structure coincides in fact with one row of the $\sqrt{5}$ surface oxide \cite{todorova03}.

\begin{figure}
\scalebox{0.75}{\includegraphics{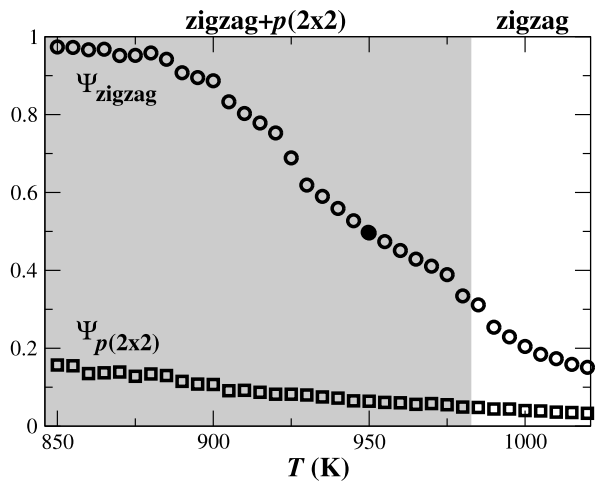}}
\caption{\label{fig3}
Order parameters $\Psi$ sensitive to the order of the zigzag step and $p(2 \times 2)$ terrace structure, as determined by grand-canonical MC simulations for $p = 10^{-3}$\,atm. At this pressure, the temperature range shown corresponds to $-1.43$\,eV $< \Delta \mu_{\rm O} < -1.16$\,eV, where the approximate thermodynamic results of Fig. \ref{fig2} predict the stability of the zigzag structure or the coexistence of zigzag + $p(2 \times 2)$ as indicated by the differently shaded backgrounds. Equivalent results are obtained for simulations at $p = 10^{-5}$ and 1 atm, with variations of the critical temperatures of the order of 100\,K.}
\end{figure}

At near atmospheric pressures the identified stability range of the zigzag step decoration ($-1.43$\,eV $< \Delta \mu_{\rm O} < -1.16$\,eV) corresponds to rather elevated temperatures of the order of 900\,K \cite{reuter01}. At such temperatures, configurational entropic effects can not be uncritically dismissed. Considering the high melting temperature of Pd and the high kink formation energy of the close-packed (111) step, we expect these effects to be dominated by a possible disorder in the adsorbate ensemble. We therefore scrutinize the insight provided by the approximate thermodynamic approach by fully accounting for such effects with grand-canonical MC simulations based on the first-principles LGH. In these simulations, the effect of the (111) step on the adsorbate ensemble in its vicinity is evaluated by separately monitoring the order and coverage for each row of sites parallel to the step edge using the same simulation cells as for the simulated annealing runs. Since the prevalent ordered structures in the targeted $(T,p)$-range are the $p(2 \times 2)$ at the terrace and the zigzag decoration at the step, order parameters that are sensitive to these lateral periodicities can be suitably employed to determine the critical temperature $T_c$ for the order-disorder transition at fixed oxygen pressure \cite{zhang07}. Interestingly, the disturbance exerted by the step on the ordering behavior on the terrace is found to be rather short ranged on the accuracy level of interest to this study. Already four atomic rows away from the immediate step edge the computed critical temperatures are to within $\pm 50$\,K identical to those obtained at the ideal Pd(100) surface. Figure \ref{fig3} therefore only contrasts the ordering behavior at the immediate step edge with that at ideal terrace sites for $p=10^{-3}$\,atm. Whereas the approximate thermodynamic approach predicts the presence of the $p(2 \times 2)$ ordered structure at the terrace up to a temperature of $980$\,K (corresponding to $\Delta \mu_{\rm O} = -1.37$\,eV at this pressure), the low order parameter obtained in the grand-canonical MC simulations shows that the terrace adlayer has in fact long disordered at these temperatures. This underscores that a full understanding of the surface structure and composition at such high temperatures can only be obtained through the explicit evaluation of the partition function as enabled by the first-principles LGH approach. In contrast, the $T_c = 950$\,K deduced from the inflection point of the order parameter for the zigzag decorated step validates that this ordered structure prevails indeed up to surprisingly high temperatures.

\section{Summary}

Through a combination of first-principles statistical mechanics techniques we have been able to identify a wide range of oxygen environments, in which (111) steps at Pd(100) will be decorated by a characteristic zigzag oxygen structure. This includes near atmospheric pressures and temperatures around 900\,K, i.e. gas-phase conditions that are representative for an important catalytic process like the high-temperature combustion of methane. Due to the low energetic cost of the close-packed steps, this defect will be a frequent structural motif at the surface with corresponding potential influence on the function in such applications. In this respect it is interesting to notice that the same ordered zigzag O structure was recently characterized at the (111) steps of Pd(111) vicinals in the same range of oxygen chemical potentials, albeit at lower temperatures \cite{westerstrom07}. Using our LAPW/APW+lo DFT setup we compute initial and final state O $1s$ core level shifts \cite{lizzit01} that are identical to within 40\,meV for the O atoms adsorbed at the geometrically equivalent Sh2 step sites in the two structures. Since core level shifts are a sensitive probe of the local bonding properties, this suggests highly similar reactivities of the oxygen atoms adsorbed at the decorated (111) steps at both prominent facets of Pd nanoparticles. If these oxygen atoms play a determining role for the catalytic function, this similarity would provide an intriguing atomic-scale interpretation for the enigmatic structure insensitivity reported recently for the methane combustion at these elevated temperatures \cite{zhu04}. While our study focuses on establishing a first methodological access to determine the structure and composition of a prominent defect like atomic steps at surfaces exposed to realistic environments, this immediately exemplifies the far-reaching insight that this kind of first-principles statistical mechanics approach can provide.

\section{Acknowledgments}

The EU is acknowledged for financial support under contract NMP3-CT-2003-505670 (NANO$_2$).


\begin{thebibliography}{99}
\bibitem{basis}
LAPW/APW+lo basis set parameters: $R_{\rm{MT}}^{\rm{Pd}}=$2.1\,bohr, $R_{\rm{MT}}^{\rm{O}}=$1.1\,bohr, $E^{\rm max}_{\rm wf} = 20$\,Ry, $E^{\rm max}_{\rm pot} = 196$\,Ry. $(12 \times 12 \times 1)$ and $(3 \times 12 \times 1)$ Monkhorst-Pack grids were used for the $(1 \times 1)$-Pd(100) and $(1 \times 1)$-Pd(117) slab calculations, respectively, and reduced in larger surface unit-cells to maintain the same reciprocal space sampling. The supercells were composed of at least 12\,{\AA} vacuum and fully relaxed 5 Pd(100) layer or 23 layer Pd(117) slabs, with adsorption of O at both sides.
\bibitem{density}
The data shown for the ordered step structures is for a step density corresponding to the $(60 \times 40)$ periodic boundary simulation cell employed to compute the LGH binding free energy. Without step-step interaction effects, the stability regions deduced from the crossings of the individual lines are independent of this specific choice of step density.
\end{thebibliography}
\end{document}